\documentclass[preprint,showpacs,preprintnumbers,amsmath,amssymb]{revtex4}

\usepackage{graphicx}
\usepackage{dcolumn}
\usepackage{bm}

\begin{document}

\title{Casimir force in brane worlds: coinciding results from Green's and Zeta function approaches}

\author{Rom\'an Linares$^1$}
\email{lirr@xanum.uam.mx}
\author{Hugo A. Morales-T\'ecotl$^1$}
\email{hugo@xanum.uam.mx}
\author{Omar Pedraza$^2$}
\email{omp@xanum.uam.mx}

\affiliation{$^1$ Departamento de F\'{\i}sica, Universidad Aut\'onoma Metropolitana Iztapalapa,\\
San Rafael Atlixco 186, C.P. 09340, M\'exico D.F., M\'exico,}

\affiliation{$^2$ Centro de Estudios en F\'{\i}sica y Matem\'aticas B\'asicas y Aplicadas, \\
Universidad Aut\'onoma de Chiapas, 4a. Oriente Norte 1428, Tuxtla
Guti\'errez, Chiapas, M\'exico,
}


\begin{abstract}
Casimir force encodes the structure of the field modes as vacuum
fluctuations and so it is sensitive to the extra dimensions of brane
worlds. Now, in flat spacetimes of arbitrary dimension the two
standard approaches to the Casimir force, Green's function and zeta
function, yield the same result, but for brane world models this was
only assumed. In this work we show both approaches yield the same
Casimir force in the case of Universal Extra Dimensions and
Randall-Sundrum scenarios with one and two branes added by $p$
compact dimensions. Essentially, the details of the mode
eigenfunctions that enter the Casimir force in the Green's function
approach get removed due to their orthogonality relations with a
measure involving the right hyper-volume of the plates and this
leaves just the contribution coming from the Zeta function approach.
The present analysis corrects previous results showing a difference
between the two approaches for the single brane Randall-Sundrum;
this was due to an erroneous hyper-volume of the plates introduced
by the authors when using the Green's function. For all the models
we discuss here, the resulting Casimir force can be neatly expressed
in terms of two four dimensional Casimir force contributions: one
for the massless mode and the other for a tower of massive modes
associated with the extra dimensions.
\end{abstract}

\maketitle
\section{Introduction}

Historically the idea to consider our observable 4D universe as a
subspace of a higher dimensional spacetime has a long tradition that
started with the works of G. N\"{o}rdstrom \cite{Nordstrom:1988fi},
T. Kaluza \cite{Kaluza:1921tu} and O. Klein \cite{Klein:1926tv} (see
e.g. \cite{Appelquist:1987nr} and references therein). Nowadays
there are two broad approaches one typically takes to address the
possible consequences of extra dimensions in 4D physics. The
top-down approach starts either from a fundamental theory or a low
energy limit of it, for instance M/string-theory  or supergravity
\cite{Polchinski:1998rq} and upon compactification of the extra
dimensions one hopes to find an effective theory in 4D containing as
much of the physics we know (see e.g. \cite{Font:2005td} and
references therein). In this approach one favors the properties of
the compactification manifold and upon the requirement that the
compactification be performed in a consistent way, one tracks the
physical consequences that the geometry of the internal manifold has
on the resulting lower dimensional theory, including, for instance, the
gauge group and the matter content. However, in this approach we are
unable to select the lower dimensional theory in a unique fashion.
The Standard Model hopefully would correspond to a particular
internal space or vacuum configuration chosen by nature by some
still unknown mechanism (see e.g. \cite{Lust:2007kw} and references
therein).

In contrast the bottom-up approach relies on ``model building'',
where the requirements of having the low energy spectrum and
interactions of the known 4D physics put restrictions on properties
such as the types of singularities, curvature, symmetries, etc.,
supported by the internal space. The constraints are powerful
because they hold for a large class of models without having to
fully specify the compactification details. Of course they are only
necessary conditions, nevertheless they serve as a useful guide in
the search for realistic models before a complete theory/model can
be explicitly constructed. In this approach one looks at the
different well known physical phenomena and their corresponding
experimental confirmations, and then, by requiring agreement between
the contributions of the extra dimensions to the 4D physics and the
experimental errors, one gets bounds to the higher dimensional free
parameters. This information forms the core of the necessary
knowledge for model building. Following this approach most attention
has been devoted to high energy physics (see e.g.
\cite{Allanach:2004ub,Csaki:2004ay} and references therein) and
cosmology (see e.g.
\cite{Maartens:2003tw,Elizalde:2003bz,Elizalde:2006iu,Maartens:2010ar}
and references therein). More recently the possibility to obtain
information from models with extra dimensions studying low energy
physical phenomena such as the Casimir effect has also been
addressed
\cite{Poppenhaeger:2003es,Linares:2005cj,Frank:2007jb,Linares:2007yz,Linares:2008am,
Frank:2008dt,Saharian:2008gs,Teo:2008ah,Teo:2009tm,Elizalde:2009nt,
Cheng:2009xf,Teo:2009dd,Cheng:2009bv,Kharlanov:2009pv,Cheng:2009uu,
Teo:2009bv,Bellucci:2009hh,Rypestol:2009pe,Nouicer:2009ze,Beneventano:2010wy,Elizalde:2010av}.
The interest in the Casimir force is twofold. Firstly, the force
between neutral perfect conducting plates predicted by H.B.G.
Casimir \cite{Casimir:1948dh}, is experimentally well established
\cite{Mohideen:1998iz,Bressi:2002fr,Decca:2005yk,Klimchitskaya:2005df,Lamoreaux:2005gf},
and nowadays the increasing accuracy reached in its determination,
makes us think that constraining model parameters in this way is at
the least complementary to those based on high energy experiments.
Secondly, its theoretical analysis involves two aspects naturally
appearing in the study of models with extra dimensions, namely the
mode structure of matter fields and the submillimeter length scale,
of order 1 $\mu$m, at which the force becomes noticeable and for
which some extra dimensional models haven been conjectured to
produce observable effects.

In this paper we follow the model building approach to determine the
Casimir force for a massless scalar field between two parallel
plates. This situation mimics the actual experimental setup where
the electromagnetic rather than a massless scalar field is
considered. We model the plates as codimension one hyper-surfaces in
the extra-dimensional space-time, therefore what one really obtains
is the force per unit of hyper-volume of the plate, as it was
established long ago for hyper-dimensional Minkowski space-time
\cite{Ambjorn:1981xw}. In this way the extra dimensions yield
corrections to the usual 4D Casimir force. Remarkably the resulting
force can be expressed as the sum of two types of contributions: one
that is given by the zero mode, thus producing the standard 4D
Casimir force for a massless scalar field, and the other one that
includes the addition of 4D Casimir forces corresponding to the
massive modes.

The present work is aimed at showing that both Green's function and
Zeta function techniques (see e.g. \cite{Elizalde:1994gf} and
references therein) yield the same Casimir force for some typical
extra dimensional scenarios. In particular it corrects a previous
difference between the Casimir force for one-brane Randall-Sundrum
models \cite{Randall:1999ee,Randall:1999vf} using the {\em zeta
function method} \cite{Frank:2007jb,Frank:2008dt} and the one
obtained using {\em Green's function approach}
\cite{Linares:2007yz,Linares:2008am}. Such difference was originated
by an erroneous hyper-volume factor for the plates considered in the
setting in \cite{Linares:2007yz,Linares:2008am}.

For the sake of clarity  we first study the case of
Universal Extra-Dimensions in 5D. This corresponds to 4D Minkowski
extended by an spatial compact extra dimension attached to each of
its points. The topology of the extra dimension is an orbifold $S^1
/ Z_2$. The Casimir effect in this geometry was studied using the
zeta function regularization method in \cite{Poppenhaeger:2003es}
but here we present the corresponding Green's function analysis. The
second model we shall consider is the so called Randall-Sundrum
II$p$ model (RSII-$p$). These have a single $(3+p)$-brane
\cite{Dubovsky:2000am}. For this model the Casimir effect was
computed in \cite{Frank:2007jb,Frank:2008dt} using the zeta function
method and in \cite{Linares:2007yz,Linares:2008am} using the Green's
function method. Finally we shall consider the Randall-Sundrum I$p$
model (RSI-$p$). These are defined by two $(3+p)$-branes. In this
case the  Casimir force was studied in \cite{Frank:2008dt} using the
zeta function technique. To the best of our knowledge, an analogous
study is missing applying the Green's function approach and in this
paper we fill in this gap. We shall conclude that for all the above
extra dimensional models the Casimir force obtained by either of the
approaches: zeta function regularization or Green's function, is the
same.

The structure of the paper is as follows. Since it will be used
frequently in Section \ref{Modes4D} we briefly recall the analysis
of the Casimir force in 4D Minkowski space-time for a massive scalar
field whereas its extension to $d+1$ Minkowski space-time is
summarized in the Appendix. In section \ref{SecUED} we discuss the
Universal Extra Dimension model. Section \ref{SecRSIIp} is devoted
to RSII-$p$ whereas Section \ref{secRSIp} deals with RSI-$p$.
Finally, Section \ref{conclusions} contains the discussion of our
results. Unless otherwise stated we use units in which $\hbar=c=1$.

\section{Scalar field in 4D Minkowski spacetime}\label{Modes4D}
To make use of it in the sequel we briefly review the analysis of
the Casimir force for a massive scalar field in 4D Minkowski
spacetime \cite{Milton:2001yy}. We start by computing the dispersion
relation and then determine the Casimir force by the two approaches:
Zeta function and Green's function.

Let the scalar field to have mass $\mu$ and subject to Dirichlet
boundary conditions at the planes $z=0,l$ . The starting point of
the analysis is the Klein-Gordon's equation
\begin{equation}\label{KG4D}
(\Box+\mu^2)\phi = 0 \,,\quad \Box = \partial_{tt} -\Delta,
\end{equation}
with $\Delta$ the Laplacian in $I\!\!R^3$ and $\mu$ the mass of the
scalar field. $x$ represents a spacetime coordinate with components
$(t,x_1,x_2,z)$, the last three being spatial and Cartesian. The
$4D$ Minkowski metric is $\eta_{\mu\nu} =
\mathrm{diag}\{1,-1,-1,-1\}$. By separating the dependence of the
field in Cartesian coordinates $\vec{x}=(x_1,x_2,z)$ as
$\phi(x_i,z,t)=\chi_i(x_i)\Upsilon(z)\, \mathrm{e}^{i\omega t}$,
$i=1,2$, we have the equivalent set of eigenvalue equations
\begin{eqnarray}
-\partial_{ii}\chi_i(x_i) &=& k_i^2\chi_i(x_i) ,\hspace{0.3cm}
i=1,2\,,\label{eq:Xi}\\
- \partial_{zz}\Upsilon(z) &=& k_3^2
\Upsilon(z) ,\hspace{0.3cm} k_3^2:= \omega^2-(\mu^2 + k_1^2 + k_2^2)
\,. \label{eq:Upsilon}
\end{eqnarray}
Here $k_i^2\,,i=1,2$, and $k_3$ are separation constants or
eigenvalues. The physical plates are 2D surfaces described by
coordinates $x_1$ and $x_2$ so that $\chi_i$ in (\ref{eq:Xi}) can be
subject to free boundary conditions. Coordinate $z$ is transverse to
the plates so $\Upsilon$ in (\ref{eq:Upsilon}) will be subject to
Dirichlet boundary conditions at $z=0,l$. The corresponding
eigenfunctions are
\begin{eqnarray}
\chi_i(x_i) &=& \frac{1}{\sqrt{2\pi}} \mathrm{e}^{ik_i
x_i},\hspace{0.3cm} k_i\in I\!\!R,\, i=1,2, \label{XiEigen}\\
\Upsilon_N (z) &=& \sqrt{\frac{2}{l}} \sin{\frac{N\pi
z}{l}},\hspace{0.3cm} N=1,2,\dots\,, \label{Upsilonn}
\end{eqnarray}
and the resulting dispersion relation is
\begin{equation}\label{DispRel}
\omega^2= \mu^2 + k_1^2 + k_2^2 + \frac{N^2\pi^2}{l^2}.
\end{equation}

\subsection{Zeta function approach}\label{DimensionalReg}

To compute the Casimir force one can compute the Casimir energy
between the plates $E_{plates}$, by summing up the zero-point energy
per unit area $\hbar \omega/2$. There are two ingredients required
to follow this strategy: the {\it dispersion relations} and {\it the
modes structure} (which can be continuous, discrete or an
admixture). It turns out that $E_{plates}$ contains a linear term in
the separation $l$ between planes which gives rise to a constant
Casimir force. This term can be canceled by addition of a constant
to the Hamiltonian density or by considering the energy $E_0$ in the
absence of the plates which means $k_3\in I\!\!R$ and $\Upsilon_N
(z)= \frac{1}{\sqrt{2\pi}} \mathrm{e}^{i\, k_3\, z}$. Adopting the
second option, the resulting finite expression for the Casimir
energy per unit area of the plate is
\begin{equation}\label{intenergydimreg}
{\mathcal E}_{4D}(\mu) = \frac{E_{plates}-E_0}{L^2} = \frac{1}{2}
\prod_{i=1,2} \int_{-\infty}^\infty \frac{dk_i}{2 \pi} \left(
\sum_{N=1}^\infty \omega_{k_1,k_2,N}(\mu) - l \int_{-\infty}^\infty
\frac{dk_3}{2 \pi} \omega_{k_1,k_2,k_3}(\mu) \right),
\end{equation}
where
\begin{equation}
\omega_{k_1,k_2,N}(\mu) \equiv \sqrt{\mu^2 + k_1^2 + k_2^2 +
\frac{N^2\pi^2}{l^2}},
\end{equation}
\begin{equation}
\omega_{k_1,k_2,k_3}(\mu) \equiv \sqrt{\mu^2 + k_1^2 + k_2^2 +
k_3^2},
\end{equation}
and $L^2$ is the area of a square shaped piece of the plates at
$z=0,l$. Notice the extra factor of $l$ in the second term of
(\ref{intenergydimreg}); it comes from the fact that $E_0$ is the
energy in the whole volume delimitated by $z=0,l$, in the transverse
direction, whereas ${\mathcal E}$ is the energy per unit area $L^2$.
We are denoting explicitly the dependence of the density energy
${\mathcal E}_{4D}$ on the mass $\mu$, to stress the fact that the
4D scalar field is massive.

We perform explicitly the integrals in the appendix \ref{apendice}
obtaining
\begin{equation}\label{energydensresult}
{\mathcal E}_{4D}(\mu) = - \frac{\mu^2}{8\pi} \cdot \frac{1}{l}
\sum_{N=1}^\infty \frac{1}{N^2}K_{-2} (2Nl\mu),
\end{equation}
where $K$ is the modified Bessel function of second type. The
Casimir force is obtained from the Casimir energy simply deriving
with respect to the separation between the plates: $F_{4D}=-\partial
{\mathcal E}_{4D}/
\partial l$, thus
\begin{equation}\label{FuerzaMassive}
f_{4D}(\mu)= - \frac{\mu^2}{8\pi^2} \left[\frac{1}{l^2}
\sum_{N=1}^{\infty}\frac{1}{N^2}K_{2}(2 Nl\mu) - \frac{2
\mu}{l}\sum_{N=1}^{\infty} \frac{1}{N}K_{3}(2 Nl\mu)\right]\,,
\end{equation}
where a property of the derivative of the Bessel function has been
used. In general this expression can not be simplified further and
usually people computes it numerically for a given value of the mass
$\mu$. For the massless case, $\mu=0$, however, the expression can
be simplified to yield
\begin{equation}
f_{4D}(0)= -\frac{\pi^2}{480} \frac{1}{l^4}\,,
\end{equation}
in which the appropriate approximation for small argument of the
Bessel functions has been used and then the identification of a zeta
function allows to evaluate the result.

\subsection{Green's function approach}\label{GreenFunctMeth}

In the Green's function approach, once we are armed with the
eigenfunctions (\ref{XiEigen}) and (\ref{Upsilonn}), we can express
the Green's function $G_{4D}$ for the problem
$(\Box+\mu^2)G_{4D}(x,x')=-\delta(x-x')$, subject to Dirichlet
boundary conditions at $z=0,l$, as
\begin{eqnarray}
G_{4D}(x,x') &=& \prod_{i=1,2}\int {dk_i}
\chi_i^{\ast}(x_i)\chi_i(x_i') \int \frac{d\omega}{2\pi}
\mathrm{e}^{-i\omega(t-t')} g(z,z'),\label{GreenFunction4D} \\
g(z,z')&=& \sum_{N=1}^{\infty} \frac{\Upsilon_N^{\ast}(z)
\Upsilon_N(z')}{\frac{N^2 \pi^2}{l^2}-k_3^2}, \label{reducedg}
\end{eqnarray}
with $\ast$ denoting complex conjugation and $g$ the so called
reduced Green's function. Notice that this expression of the Green
function is valid only in the region between the plates. Now given
the relation between the vacuum expectation value of the time
ordered product of fields and the Green's function, $\langle
T[\phi(x)\phi(x')]\rangle =\frac{1}{i} G_{4D}(x,x')$, the force per
unit area on either plate can be obtained from the vacuum
expectation value of the energy-momentum tensor $T_{\nu
\rho}=\partial_{\nu}\phi\partial_{\rho}\phi-\eta_{\nu \rho}{\cal L}$
with ${\cal L}=
\frac{1}{2}\partial_{\nu}\phi\partial^{\nu}\phi-\frac{1}{2}\mu^2\phi^2$.
Since upon integration over all space the term $\eta_{\nu\rho}{\cal
L}$ does not contribute by virtue of the Klein-Gordon's equation
(\ref{KG4D}) one gets
\begin{eqnarray}
f_{in}(\mu) &=& \langle T^{in}_{zz}\rangle=
\left.\lim_{x'\rightarrow
x} \partial_z\partial_{z'}G_{4D}(x,x')\right|_{z=0,l}\\
&=& \prod_{i=1,2}\int {dk_i} \chi_i^{\ast}(x_i)\chi_i(x_i)
\int\frac{d\omega}{2\pi} \left.\lim_{z'\rightarrow z}
\partial_z\partial_{z'}g(z,z')\right|_{z=0,l} \label{limitg}
\end{eqnarray}
In the coincident limit
$\chi_i^{\ast}(x_i)\chi_i(x_i)=\frac{1}{2\pi}$ and so the dependence
on $x_i,i=1,2,$ drops out; which should have been expected from the
translational invariance of the parallel plates configuration along
the $x_i, i=1,2$, directions.

Combining (\ref{Upsilonn}) together with (\ref{reducedg}) allows us
to obtain the explicit form of $g(z,z')$, which upon substitution in
(\ref{limitg}) and after the change of variables: $\omega\rightarrow
i\xi$ and $k_1^2+k_2^2+\xi^2 \rightarrow \rho^2$ produces
\begin{eqnarray}\label{IntDiverGreen}
f_{in}(\mu) &=& \frac{1}{l} \int \frac{dk_1dk_2}{(2\pi)^2} \int
\frac{d\xi}{2\pi} \sum_{N=1}^{\infty} \frac{\frac{\pi^2
N^2}{l^2}}{\rho^2 + \mu^2 + \frac{\pi^2 N^2}{l^2}}.
\end{eqnarray}
This allows us to read $(k_i,\xi)$ as
three-dimensional Cartesian coordinates. This is not the final
answer because so far we have only considered the force to the left
of $z=l$ or to the right of $z=0$; actually the integral
(\ref{IntDiverGreen}) diverges. We also have to include the flux of
momentum for instance to the right of $z=l$. We shall not elaborate
on this issue, for our purpose it is enough to mention that the
normal-normal component of the stress tensor at $z=l$ is: $\langle
T^{out}_{zz}\rangle=i \sqrt{\rho^2+\mu^2} /2$ \cite{Milton:2001yy}.
The net forces at $z=l$ produces finally
\begin{eqnarray}\label{eq:f4Dmu0}
f_{4D}(\mu) &=& \frac{1}{2} \int \frac{dk_1dk_2}{(2\pi)^2} \int
\frac{d\xi}{2\pi} \left(  \frac 2l \sum_{N=1}^{\infty}
\frac{\frac{\pi^2 N^2}{l^2}}{\rho^2 + \mu^2 + \frac{\pi^2
N^2}{l^2}}+ \sqrt{\rho^2+\mu^2} \right).
\end{eqnarray}
We compute this integral explicitly in the appendix \ref{apendice},
obtaining again the expression (\ref{FuerzaMassive}) for the 4D
Casimir force. We could use the easier argument based on the
discontinuity of the derivative of the reduced Green's function $g$
above. Of course the same $f_{4D}$ results from the discontinuity of
the $zz$ component of the energy momentum tensor on either plate at
$z=0,l$.

Notice that in order to obtain the force, the Green's function
method includes a ``bit" more of information than the previously
used in the zeta function regularization method. We have used again
the eigenvalues through the {\it dispersion relations} but we have
also used explicitly the {\it eigenfunctions} and not only the mode
structure. However upon integration of the modes $\chi_i$, $i=1,2$,
the real input is again only the modes structure as in the zeta
regularization method loosing the ``bit" of extra information. This
property of the Green's method is the one that makes it equivalent
to the zeta function regularization method. For instance in the
models with extra dimensions, as we will discuss, the eigenfunctions
depending on the extra coordinates give us information about the
localization of the field modes, however, since at the end these
eigenfunctions are integrated out, the information on the
localization is in some sense ``lost". This is in agreement with our
concept of consistent compactifications, for which the extra
dimensional coordinates must disappear explicitly (see for instance
\cite{Duff:1986hr} and references therein).

It is straightforward to generalize the result of the 4D Casimir
force to a $(d+2)$D Minkowski spacetime (see appendix
\ref{apendice}). In this case the scalar field is bound by
hyperplanes of $d$ dimensions and the force per unit $d$-dimensional
volume between the hyperplanes is given by \cite{Ambjorn:1981xw}
\begin{equation}
f_{(d+2)D}(\mu)= -2 \left
(\frac{\mu}{4\pi}\right)^{\frac{d+2}{2}}\left[\frac{3}{l^{\frac{d+2}{2}}}
\sum_{N=1}^{\infty}\frac{1}{N^{\frac{d+2}{2}}}K_{\frac{d+2}{2}}(2
Nl\mu) + \frac{2 \mu}{l^{\frac{d}{2}}}\sum_{N=1}^{\infty}
\frac{1}{N^{\frac{d}{2}}}K_{\frac{d}{2}}(2 Nl\mu)\right].
\end{equation}
For the forthcoming analysis it is convenient to notice that this
expression has the following limit values
\begin{equation}
\lim_{\mu \rightarrow 0 }f_{(d+2)D}(\mu)= -\frac{d}{l^{d+2}(4
\pi)^{\frac{d+2}{2}}} \Gamma\left( \frac{d+2}{2} \right) \zeta(d+2),
\end{equation}
and
\begin{equation}
\lim_{\mu \rightarrow \infty } f_{(d+2)D}(\mu) \rightarrow 0.
\end{equation}
Once we have reviewed the way in which each method gives origin to
the Casimir force, let us continue with the extra dimensions models.

\section{Universal Extra Dimensions}\label{SecUED}

\subsection{The model}\label{UEDModel}

The Casimir force in a Universal Extra Dimension (UXD) scenario
\cite{Appelquist:2000nn} was considered in
\cite{Poppenhaeger:2003es} for the case of a massless scalar field
to probe the possible existence and size of an additional spatial
dimension which is compactified on a $S^1/Z_2$ orbifold. This
geometry restricts the possible vacuum fluctuations of the scalar
field to have a wave vector along the extra dimension of the form
$k_n=n/R$, with $k_n$ being the wave vectors in the direction of the
universal extra dimension and $R$ the radius of $S^1$.

Let us start with the 5D action for a massive scalar field
\begin{equation}
S= \frac12 \int d^4x \int_0^{\pi}  R \, d \theta \, \sqrt{g} \left(
g^{\alpha \beta} \partial_\alpha \Phi(x, \theta)
\partial_\beta \Phi(x, \theta)-m_5^2 \Phi^2(x,\theta)\right).
\end{equation}
Here $x^\alpha=(x^\mu, R \theta)$ are the coordinates with $\alpha
=(\mu, 4)$ and $\mu=0, \dots ,3$ are the indexes of our 4D
spacetime. $m_5$ is the mass of the 5D field. $\pi R$ is the size of
the extra dimensions and $g^{\alpha \beta}$ is the inverse of the
metric defined by the interval
\begin{equation}\label{mebra}
ds_{5}^{\, 2}=\eta_{\mu\nu}dx^{\mu}dx^{\nu}-R^2d \theta^2.
\end{equation}
In this metric, the 5D Klein-Gordon equation reads
\begin{equation}
\Box_4 \Phi-\frac{1}{R^2} \partial_{\theta}^2 \Phi + m_5^2 \Phi=0,
\end{equation}
which separates through $\Phi(x,R \theta)= \phi(x) \psi(\theta)$
into
\begin{eqnarray}
\left( \partial_{\theta}^2+m_{\theta}^2R^2  \right)  \psi(\theta)
&=&0,\\
\Box_4 \phi+ \left( m_5^2+ m_{\theta}^2 \right) \phi(x)&=&0.
\end{eqnarray}
When the extra dimension is Kaluza-Klein type, the only condition on
the fields is: $\Phi(x, \theta)= \Phi(x, \theta+2\pi)$, which allows
for a Fourier expansion taking the form
\begin{equation}
\Phi(x, \theta)=\frac{1}{\sqrt{\pi R}} \phi^{(0)}(x) +
\sum_{n=1}^\infty \sqrt{\frac{2}{\pi R}} \left[\phi^{(n)}(x) \cos
(n\theta)+ \chi^{(n)}(x) \sin (n \theta) \right].
\end{equation}
When the extra dimension is instead an orbifold $S^1/Z_2$, there is
an additional parity condition on the 5D scalar field: $\Phi(x,
\theta)= \pm \Phi(x, -\theta)$. In this case it is clear that the
modes of the scalar field have definite parity, they are either
even, denoted by (+), or odd, denoted by (-). Explicitly, they are
\begin{eqnarray}
\phi(x,\theta)_{+}&=&\frac{1}{\sqrt{\pi R}}\,
\phi^{(0)}(x)+\sum_{n=1}^{\infty}\sqrt{\frac{2}{\pi
R}}\, \phi^{(n)}(x)\cos (n \theta ),\\
\phi(x,\theta)_{-}&=&\sum_{n=1}^{\infty}\sqrt{\frac{2}{\pi R}}\,
\chi^{(n)}(x)\sin (n \theta).
\end{eqnarray}
In models of extra dimensions, the zero mode is associated with the
lower dimensional physics, and, as a consequence, we need to
consider the even modes if we want to reproduce the lower
dimensional physics. Therefore we impose the additional parity
condition $\Phi(x, -\theta)=+ \Phi(x, \theta)$ on the 5D scalar
field which leave us with the set of 4D scalars field $\{
\phi^{(0)}(x) , \phi^{(n)}(x) \}$. These fields satisfy the
effective 4D equations
\begin{eqnarray}
\Box_4 \phi^{(0)}(x)+ m_5^2 \phi^{(0)}(x)&=&0, \\
\Box_4 \phi^{(n)}(x)+ \left( m_5^2+ \frac{n^2}{R^2} \right)
\phi^{(n)}(x)&=&0.
\end{eqnarray}
We interpret to the zero mode $\phi^{(0)}(x)$ as a 4D massive scalar
field of mass $m_5$ and to each mode of the Kaluza-Klein tower
$\phi^{(n)}(x)$ as 4D massive scalar fields of mass $m_4$, given by
\begin{equation}\label{m4UED}
m_4 \equiv \sqrt{m_5^2+ \frac{n^2}{R^2}}, \hspace{1cm} n \in
\mathbb{N} \cup \{0 \}.
\end{equation}
The modes for the coordinate $\theta$ are simply
\begin{equation}
\psi_{0}(\theta)=\frac{1}{\sqrt{\pi R}}\hspace{0.5cm}
\mbox{and}\hspace{0.5cm} \psi_{n}(\theta)=\sqrt{\frac{2}{\pi R}}\cos
(n \theta ).\label{ecdmndc1}
\end{equation}
This means that the zero mode field is constant along the extra
dimension whereas the massive modes are distributed harmonically in
that direction.

\subsection{Zeta function approach}\label{UEDDimReg}

In \cite{Poppenhaeger:2003es} the zeta function
method was used to compute the Casimir force associated to a 5D
massless scalar field. Here we consider a 5D massive scalar field.
As we have mentioned, in this formalism the relevant quantities are
the frequency of the vacuum fluctuations and the modes structure, so
if we decompose the 4D fields (i.e. the zero mode and the
Kaluza-Klein tower) in the same way as in equations (\ref{eq:Xi}) to
(\ref{Upsilonn}), the energy per unit 3-volume ($L^2 \times \pi R$)
of the hyperplanes is
\begin{equation}\label{EUXDm5}
{\mathcal E}_{UXD}(m_5)= \frac{1}{2} \prod_{i=1,2}
\int_{-\infty}^\infty \frac{dk_i}{2 \pi}  \left(
\sum_{N=1,n=0}^\infty \omega_{k_i,N,n}\left(m_4\right)- l
\int_{-\infty}^\infty \frac{dk_3}{2 \pi} \sum_{n=0}^\infty
\omega_{k_i,k_3,n}\left(m_4 \right) \right) \, ,
\end{equation}
where
\begin{equation}
\omega_{k_i,N,n}\left(m_4 \right) \equiv \sqrt{k_{1}^2+ k_{2}^2
+\left(\frac{\pi N}{l}\right)^2+\frac{n^2}{R^2}+m_5^2},
\end{equation}
\begin{equation}
\omega_{k_i,k_3,n}\left( m_4 \right) \equiv \sqrt{k_{1}^2+ k_{2}^2 +
k_3^2 + \frac{n^2}{R^2}+m_5^2}.
\end{equation}
and the mass $m_4$ is given by (\ref{m4UED}).

In order to get ${\mathcal E}_{UXD}$ we do not have to compute
anything, we already have the answer. Notice that we can rewrite Eq.
(\ref{EUXDm5}) as follows
\begin{equation} {\mathcal E}_{UXD}(m_5)=
{\mathcal E}_{4D}(m_4 |_{n=0} )+ \sum_{n=1}^\infty {\mathcal
E}_{4D}\left(m_4 \right),
\end{equation}
where ${\mathcal E}_{4D}(m_5)$ is given by (\ref{intenergydimreg})
and whose analytical expression after integration is
(\ref{energydensresult}). Deriving this expression with respect to
the separation between hyperplanes we finally get
\begin{equation}\label{forceUXDfinalshort}
f_{UXD}(m_5)= f_{4D}(m_5)+ \sum_{n=1}^\infty f_{4D}\left(m_4
\right),
\end{equation}
which reads
\begin{eqnarray}
f_{UXD}(m_5)&=& - \frac{m_5^2}{8\pi^2} \left[\frac{3}{l^2}
\sum_{N=1}^{\infty}\frac{1}{N^2}K_{2}(2 Nlm_5) + \frac{2
m_5}{l}\sum_{N=1}^{\infty} \frac{1}{N}K_{1}(2 Nl m_5)\right]
\nonumber \\
& & - \frac{1}{8\pi^2} \sum_{n=1}^\infty
\left(m_5^2+\frac{n^2}{R^2}\right) \left[\frac{3}{l^2}
\sum_{N=1}^{\infty}\frac{1}{N^2}K_{2}\left( 2
Nl\sqrt{m_5^2+\frac{n^2}{R^2}}\right) \right. \nonumber \\
& & \hspace{5cm} \left. + \frac{2
\sqrt{m_5^2+\frac{n^2}{R^2}}}{l}\sum_{N=1}^{\infty}
\frac{1}{N}K_{1}\left(2
Nl\sqrt{m_5^2+\frac{n^2}{R^2}}\right)\right]. \nonumber
\end{eqnarray}
For the massless case ($m_5=0$) \cite{Poppenhaeger:2003es}, one
obtains
\begin{eqnarray}
f_{UXD}(m_5=0)&=& - \frac{\pi^2}{480} \frac{1}{l^4}+
\sum_{n=1}^\infty
f_{4D}\left(\frac{n}{R}\right) \\
&=& - \frac{\pi^2}{480} \frac{1}{l^4}- \frac{1}{8\pi^2}
\sum_{n=1}^\infty \frac{n^2}{R^2} \left[\frac{3}{l^2}
\sum_{N=1}^{\infty}\frac{1}{N^2}K_{2}\left( 2 Nl \frac{n}{R}\right)
+ \frac{2n}{l R} \sum_{N=1}^{\infty} \frac{1}{N}K_{1}\left(2 Nl
\frac{n}{R}\right)\right]. \nonumber
\end{eqnarray}
In the setting with only one extra dimension, a good agreement with
the data can only be obtained if the radius of such dimension is
smaller than $R \leq$10nm \cite{Poppenhaeger:2003es}. This bound is
weaker than others obtained from high energy physics which are
around $10^{-9}$nm.

\subsection{Green's function approach}\label{UEDGreenFunc}

Consider now the 5D Green's function expressed in terms of the
eigenfunctions (\ref{ecdmndc1})
\begin{equation}
G_{5D}(x,\theta,x',\theta')= \sum_{n=0}^\infty \psi_n(\theta)
\psi_n(\theta ') G_{4D}\left(x,x'; m_5^2+\frac{n^2}{R^2}\right),
\end{equation}
where $G_{4D}$ is the 4D Green's function given by
(\ref{GreenFunction4D}). In terms of it, the Casimir force between
the hyperplanes is
\begin{equation}\label{forceUXDtermsT}
f_{UXD}= \frac{1}{L^2} \int d \vec{x}_{\perp} \int_{0}^\pi R d\theta
\sqrt{g} \left[ \langle T_{zz}^{in}\rangle|_{z=l} - \langle
T_{zz}^{out}\rangle|_{z=l} \right],
\end{equation}
where
\begin{equation}\label{expectationT5D}
\langle T_{zz}^{in/out}\rangle|_{z=l}= \frac{1}{2i} \partial_z
\partial_{z'}G_{5D}^{in/out}(x,\theta,x',\theta')|_{x_i \rightarrow x_i',\theta \rightarrow
\theta'}.
\end{equation}
Just as with the zeta function method, we do not have to compute
that much to get the answer. Notice that the 5D Green's function can
be rewritten in terms of the 4D one in the way
\begin{equation}
G_{5D}(x,\theta,x',\theta')= \psi_0(\theta) \psi_0(\theta ')
G_{4D}\left(x,x'; m_5^2\right)+ \sum_{n=1}^\infty \psi_n(\theta)
\psi_n(\theta ') G_{4D}\left(x,x'; m_5^2+\frac{n^2}{R^2}\right),
\end{equation}
and the expectation value (\ref{expectationT5D}) can
be rewritten in terms of the expectation values in 4D
\begin{equation}
\langle T_{zz}^{in/out}\rangle|_{z=l}=  \psi^2_0(\theta)
\frac{1}{2i} \partial_z
\partial_{z'}G_{4D}^{in/out}(x,x')|_{x_i \rightarrow x_i'} +
\sum_{n=1}^\infty \psi^2_n(\theta) \frac{1}{2i} \partial_z
\partial_{z'}G_{4D}^{in/out}(x,x')|_{x_i \rightarrow x_i'}.
\end{equation}
Substituting this expression in (\ref{forceUXDtermsT}) we can
rewrite the force in terms of the Casimir force for scalars fields
in 4D
\begin{equation} \label{f-UEXD-G}
f_{UXD}(m_5)= f_{4D}(m_5) \int_0^\pi  R d \theta \, \psi^2_0(\theta)
+ \sum_{n=1}^\infty f_{4D}\left(m_5^2+\frac{n^2}{R^2} \right)
\int_0^\pi R d \theta \, \psi^2_n(\theta).
\end{equation}
Because both integrals in $\theta$ are equal to 1, we get exactly
the expression (\ref{forceUXDfinalshort}). We conclude that in the
case of one Universal Extra Dimension, the effective Casimir forces
obtained by both methods coincide. Notice that he force we have
computed is the force per unit volume, i.e. the force per unit area
($L^2$) of the plates and per unit length in the extra dimension.
Physically what we have is a couple of 3D plates, with one dimension
stretching along the extra dimension, but both embedded in four
spatial dimensions. Such a setting is referred to as having plates
of codimension one. An extension of this idea to Randall-Sundrum
models actually holds and is what we show next.

\section{Randall-Sundrum II-$p$ models}\label{SecRSIIp}

The interest in the Randall-Sundrum II-$p$ models comes from its
property of localizing not only scalar and gravity fields but also
gauge fields whenever there are $p$ extra compact dimensions
\cite{Dubovsky:2000am,Dubovsky:2000av,Oda:2000zc}. In the case of
$p=0$ the model only localizes scalar and gravity fields. The model
corresponds to a $(3+p)$-brane with $p$ compact dimensions and
positive tension $\kappa$, embedded in a $(5+p)$ spacetime whose
metrics are two patches of anti-de Sitter (AdS$_{5+p}$) of curvature
radius $\kappa^{-1}$
\begin{equation}\label{mebra}
ds_{5+p}^{\, 2}=e^{-2\kappa
|y|}\left[\eta_{\mu\nu}dx^{\mu}dx^{\nu}-\sum_{j\,=1}^pR^{\,
2}_jd\theta_j\right]-dy^2\,.
\end{equation}
The Casimir force for a massless scalar field in the RSII setup
($p=0$) was computed in \cite{Frank:2007jb} using the zeta function
regularization method, whereas in \cite{Linares:2007yz} the Casimir
force was computed for both a massive and a massless scalar field in
the RSII-1 model and then generalized to the RSII-$p$ model by means
of the Green's function approach in \cite{Linares:2008am} and using
the zeta function method in \cite{Frank:2008dt}. However their
results turned out different and seemingly depended on the method
adopted - a situation clearly unacceptable. Here we will show that
such difference was originated by an erroneous hyper-volume factor
for the plates considered in the setting in
\cite{Linares:2007yz,Linares:2008am}. We shall restrict ourselves to
the case of a higher dimensional massless scalar field. The
interested reader in the massive case can see \cite{Linares:2008am}
performing the corresponding modifications. A related calculation of
the Casimir effect in de Sitter and anti-de Sitter braneworlds can
be found in \cite{Elizalde:2002dd}.

\subsection{The mode structure}

Let us consider the $(5+p)$D action for a
massless scalar field $\Phi$ in the RSII-$p$ metric (\ref{mebra})
\begin{equation}\label{action}
S=\frac{1}{2} \int \, d{}\,^4x  \,  \prod_{j=1}^p \, R_j d \theta_j
\, dy\, \sqrt{|g|} \, g^{\alpha \beta}\partial_\alpha \Phi \,
\partial_\beta \Phi .
\end{equation}
Here $X^\alpha \equiv(x^\mu,R_i \theta_i, y)$ where $x^\mu$ are the
coordinates of our 4D spacetime, the $p$ coordinates $\theta_i$ are
associated to the $p$ compact $S^1$'s and $y$ is the noncompact
coordinate transverse to the brane which is placed at $y=0$. The
field equation for the scalar field is given by
\begin{equation}\label{EqOfMotion}
e^{2\kappa|y|}\Box_4\Phi-e^{2\kappa|y|}\sum_{j=1}^p\frac{1}{R_j^{\,
2}} \, \partial_{\theta_j}^{\, 2}\Phi -\frac{1}{\sqrt{-g}}\,
\partial_{y}\left[\sqrt{-g} \partial_{y}\Phi\right]=0,
\end{equation}
which separates through $\Phi(X)=\varphi(x) \prod_{j=1}^p\,
\Theta_j(\theta_j)\psi(y)$ into
\begin{eqnarray}
\left(\partial_{\theta_j}^{\, 2}+ m_{\theta_j}^{\,2}
R_j^{\,2}\right) \Theta_j (\theta_j)& = & 0,\qquad
j=1,\dots,p, \label{Thetaeqn} \\
\left(\partial_{y}^{2}-(4+p)\kappa\,sgn(y)\partial_{y} +
m^2\,e^{2\kappa|y|}\right)\psi(y) & = & 0, \label{Yeqn}\\
\left(\Box_4 + m_4^{2} \right)\varphi
(x)& = & 0.\label{varphieqn}
\end{eqnarray}
The  $(p+1)$ separation constants with units of mass, $m_{\theta_j}$
and  $m$, correspond to the spectra of the modes for the compact and
non compact dimensions, respectively. They give rise in turn to the
effective mass, $m_4$,  of the 4D modes in (\ref{varphieqn})
through: $ m_4^2 \equiv \sum_{j=1}^p \, m_{\theta_j}^{\,2}+m^2 $.

To find mode solutions to the above equations we shall incorporate
three types of boundary conditions: (a) To implement the presence of
the plates in ($4+p$)-space we simply set $\varphi(z=0,l)=0$. The
eigenfunctions and eigenvalues for this Dirichlet boundary
conditions were already discussed in section \ref{Modes4D}. (b) To
match the modes across the brane along the non compact dimension we
impose $\psi(y=0^{+})=\psi(y=0^{-})$ and
$\partial_y\psi(y=0^{+})=\partial_y\psi(y=0^{-})$. (c) To account
for the compactness of the $p$ dimensions we set
$\Theta_{n_j}(\theta_j)=\Theta_{n_j}(\theta_j+2\pi)$. Hereby we
obtain explicitly the plates represented by two parallel planes in
3-space but stretching along the extra dimensions.

The allowed modes for the non compact dimension are now a massless
zero mode localized on the brane
\begin{equation}\label{zeromode}
\psi_0=\sqrt{\frac{(2+p)\kappa}{2}},
\end{equation}
which satisfies the normalization condition,
\begin{equation}\label{OrtogonalityRelationsIIP}
\int_{-\infty}^{\infty} \, dy \,  e^{-(p+2)\kappa |y|}\, \psi_0^2
=1.
\end{equation}
The localization comes from the fact that the 4D
effective profile of the modes is given by $\tilde{\psi}_0=
e^{-(p+2)\kappa |y|/2}\, \psi_0$ which clearly is localized on the
brane. The massive modes have the form
\begin{equation}\label{psi}
\psi_m(y)=e^{\frac{4+p}{2}\kappa y}\,\sqrt{\frac{m}{2\kappa}}\left[
a_m J_{\gamma}\left(\frac{me^{\kappa y}}{\kappa}\right)+ b_m
N_{\gamma}\left(\frac{me^{\kappa y}}{\kappa}\right)  \right], \qquad
m>0.
\end{equation}
Here $J_\gamma$ and $N_\gamma$ are the Bessel and Neumann functions
respectively. $\gamma=\frac{4+p}{2}$ and the coefficients $a_m$ and
$b_m$ are given by
\begin{equation}\label{coeficients}
a_m=-\frac{A_m}{\sqrt{1+A_m^2}},\quad b_m=\frac{1}{\sqrt{1+A_m^2}},
\end{equation}
where
\begin{equation}\label{Aexpression}
A_m=\frac{N_{\gamma-1}\left(\frac{m}{\kappa}\right)}{
J_{\gamma-1}\left(\frac{m}{\kappa}\right)}.
\end{equation}
Notice that in this case the localization of the massive modes on
the brane is better for increasing $p$, since the modes are
modulated exponentially in the form $e^{-p \kappa |y|/2}$. The
normalization condition for the massive modes is
\begin{equation} \label{RSIIp-psi-mass}
\int_{-\infty}^\infty dy e^{-(p+2)\kappa |y|} \psi_m(y) \psi_{m'}(y)=
\delta(m-m').
\end{equation}
The modes in $\theta_j$ are:
\begin{equation}\label{Theta}
\Theta_{n_j}(\theta_j)=\frac{1}{\sqrt{2\pi R_j}}e^{i\, n_j\,
\theta_j}\hspace{1cm} \mbox{where} \hspace{1cm} n_j=m_{\theta_j} R_j
\in {\mathbb{Z}}.
\end{equation}
Therefore the contributions of the extra compact dimensions to
$m_4$, are given in terms of $m_{\theta_j}^{\,2}=n_j^2/R_j^2$
\begin{equation}\label{m4RSIIp}
m_4^2 = \sum_{j=1}^p \frac{n_j^2}{R_j^2} + m^2.
\end{equation}

\subsection{Zeta function approach}

In analogy with the cases studied above and due to the fact that the
modes behave differently for the zero mode (\ref{zeromode}) and for
the KK modes (\ref{psi}), the energy density per unit of ($p+3$) volume
$ \left( L^2 \times \prod_{j=1}^p (2\pi R_j) \times \frac{2}{(p+3)
\kappa} \right)$ for the scalar field is
\begin{eqnarray}
{\mathcal E}_{RSIIp} &=& \frac{1}{2} \prod_{i=1,2}
\int_{-\infty}^\infty \frac{dk_i}{2 \pi} \sum_{\{n\}} \left(
\sum_{N=1}^\infty \omega_{k_i,N,n_j}\left( m_4 \right)- l
\int_{-\infty}^\infty \frac{dk_3}{2 \pi}
\omega_{k_i,k_3,n_j}\left(m_4 \right)
\right) \\
&+&  \frac{1}{2} \prod_{i=1,2} \int_{-\infty}^\infty \frac{dk_i}{2
\pi} \int_0^\infty \frac{dm}{\kappa} \sum_{\{n\}} \left(
\sum_{N=1}^\infty \omega_{k_i,N,n_j,m}\left(m_4 \right)-  l
\int_{-\infty}^\infty \frac{dk_3}{2 \pi}
\omega_{k_i,k_3,n_j,m}\left(m_4 \right) \right) \nonumber
\end{eqnarray}
where $m_4$ is given by (\ref{m4RSIIp}), $\{n\}$ denotes the set
$\{n_1, n_2, \dots, n_p|n_1 \in {\mathbb{Z}}, \dots , n_p \in
{\mathbb{Z}} \}$ and the dispersion relations are
\begin{equation}
\omega_{k_i,N,n_j,m}\left(m_4 \right) \equiv \sqrt{k_{1}^2+ k_{2}^2
+\left(\frac{\pi
N}{l}\right)^2+\sum_{j=1}^{p}\frac{n_j^2}{R_j^2}+m^2},
\end{equation}
and
\begin{equation}
\omega_{k_i,k_3,n_j,m}\left(m_4 \right) \equiv \sqrt{k_{1}^2+
k_{2}^2 + k_3^2 + \sum_{j=1}^{p} \frac{n_j^2}{R_j^2}+m^2}.
\end{equation}

It is important to stress that in this case each one of the
different $p$ sums in $n_j$ goes from $-\infty$ to $\infty$ and not
as in the UXD case from $0$ to $\infty$. The reason is that here we
are considering the Kaluza-Klein tower associated to $S^1$ whereas
in the UXD case the tower is due to the orbifold $S^1/Z_2$. In fact
the Kaluza-Klein tower of $S^1$ is two copies the Kaluza-Klein tower
of $S^1/Z_2$.

Regarding the integration on the continuous massive modes $m$ it is
possible to take advantage of the following trick
\begin{equation}\label{trick}
\int_0^\infty dm f(m)= \frac{1}{2} \int_{-\infty}^\infty dm f(m) =
\pi \int_{-\infty}^{\infty} \frac{dm}{2 \pi} f(m),
\end{equation}
which is valid whenever the function $f$ be even: $f(-m)=f(m)$. Due
to the fact that the frequency $\omega(m)$ satisfies this condition,
one can consider the integration on $m$ at the same footing that the
integrals on $k_i$, $i=1,2$. As a consequence after integration, the
result is simply
\begin{equation}
{\mathcal E}_{RSIIp}(0)= \sum_{\{n\}} \left( {\mathcal
E}_{4D}\left(\sqrt{\sum_{j=1}^{p} \frac{n_j^2}{R^2}}\right) +
\frac{\pi}{\kappa} {\mathcal E}_{5D}\left(\sqrt{\sum_{j=1}^{p}
\frac{n_j^2}{R^2}}\right)\right),
\end{equation}
where ${\mathcal E}_{5D}(\mu)$ can be obtained from
(\ref{energyint}) and whose analytical expression after integration
can be obtained from (\ref{EnergyGenRes}). Deriving this expression
with respect to the separation between hyperplanes one gets
\begin{equation}\label{forceRSIIfinalshort}
f_{RSIIp}(0)=  \sum_{\{n\}} \left( f_{4D}\left(\sqrt{\sum_{j=1}^{p}
\frac{n_j^2}{R^2}}\right) + \frac{\pi}{\kappa}
f_{5D}\left(\sqrt{\sum_{j=1}^{p} \frac{n_j^2}{R^2}}\right)\right).
\end{equation}
Interpretation of this result is straightforward, the first term
corresponds to an infinite sum of 4D Casimir forces, one of the
terms corresponds to the Casimir force due to a massless scalar
field (the one corresponding to the zero mode of all the $p$ compact
extra dimensions), plus an infinite sum of 4D Casimir forces
corresponding to massive scalar fields (where the mass corresponds
to all different combinations where there is at least a non zero
mode), the second term comes from the non zero modes of the non
compact extra dimensions and corresponds to a sum of 5D Casimir
forces. Some examples previously discussed in the literature are:

\noindent $\bullet$ Case $p=0$.

In this case we simply have \cite{Frank:2007jb}
\begin{equation}
f_{RSII}(0)=  f_{4D}(0) + \frac{\pi}{\kappa} f_{5D}(0).
\end{equation}
At this point you can think units do not match, but they do because
$f_4$ is a force per unit area and $f_5$ is a force per unit volume.
Explicitly we have
\begin{equation}
f_{RSII}(0)= - \frac{\pi^2}{480} \frac{1}{l^4} \left(1+\frac{45}{4
\pi^3} \zeta(5) \frac{1}{\kappa l} \right).
\end{equation}

\noindent $\bullet$ Case $p=1$.

In this case we have \cite{Frank:2008dt}
\begin{eqnarray}
f_{RSII1}(0) &=&  \sum_{n=-\infty}^\infty \left(
f_{4D}\left(\frac{n}{R}\right) + \frac{\pi}{\kappa} f_{5D}\left(
\frac{n}{R} \right)\right) \nonumber \\
&=&   f_{4D} (0) + \frac{\pi}{\kappa} f_{5D}(0) + 2
\sum_{n=0}^\infty \left( f_{4D}\left(\frac{n}{R}\right) + \frac{
\pi}{\kappa} f_{5D}\left( \frac{n}{R} \right)\right)
\end{eqnarray}

\subsection{Green's function approach}

Lets now apply the Green's function method to the RSII$p$ models. As
we have discussed the force between the plates is obtained by
integrating over coordinates ``lateral" to the plates. In this case:
$\vec{x}_{\perp},y,\theta_j$ due to the fact that the normal-normal
component of vacuum energy momentum tensor in 3+1+$p$ spatial
dimensions has physical units of force per unit of ``volume" of
2+1+$p$ space:
\begin{equation}\label{ForceT}
F = \int_0^A d\vec{x}_{\perp} \int_{-\infty}^{\infty} dy
\sqrt{|g_{\mathrm{plate}}|} \left[\prod_{j=1}^{p}\int_0^{2\pi}
Rd\theta_j\right] \left[\langle
T_{zz}^{\mathrm{in}}\rangle\bigg|_{z=l} - \langle
T_{zz}^{\mathrm{out}}\rangle\bigg|_{z=l}\right],
\end{equation}
where $A$ is the area of the planes forming the plates in 3-space
and $g_{\mathrm{plates}}$ is the induced metric on the physical
plate located a $z=l$. Since the physical plate is a surface of
$(p+2)$D, the $\sqrt{g_{\mathrm{plates}}}$ contributes an
exponential of $-\kappa | y | (p+2)$. Instead of this factor the
exponential in \cite{Linares:2007yz,Linares:2008am} contained an
erroneous power given by  $-\kappa | y | (p+3)$.

The vacuum expectation values and the Green's function are related
to the normal-normal components of the vacuum energy momentum tensor
through
\begin{eqnarray}\label{t2}
\langle T_{zz}^{\mathrm{in/out}}\rangle\bigg|_{z=l} &=& \frac{1}{2i}
\partial_z\partial_{z'}G_{(5+p)D}^{\mathrm{in/out}}(x,y,\theta;x',y',\theta
')\bigg|_{x_\perp \rightarrow x_\perp', \, z\rightarrow z'=l,\,
\theta_j\rightarrow\theta_j '} \,.
\end{eqnarray}
As in the previous cases we can rewrite the ($5+p$)D Green's
function in terms of the 4D Green's function
\begin{eqnarray}
G_{(5+p)D}(x,y,\theta;x',y',\theta ') &=& \sum_{\{n \}}
\prod_{j=1}^p \Theta_{n_j}^*(\theta_j) \Theta_{n_j}(\theta'_j)
\psi_0(y) \psi_0(y') G_{4D}(x,x'; m_4|_{m=0}) \nonumber \\
&+& \sum_{\{n \}} \int \frac{dm}{\kappa} \prod_{j=1}^p \Theta_{n_j}^*(\theta_j)
\Theta_{n_j}(\theta'_j) \psi_m(y) \psi_m(y') G_{4D}(x,x'; m_4),
\end{eqnarray}
where $m_4$ is given by (\ref{m4RSIIp}). Introducing this expression
in (\ref{t2}) and then in (\ref{ForceT}) we obtain
\begin{equation}\label{f-RSIIp-G}
f= \sum_{\{n\}} \, f_{4D}(m_4|_{m=0}) \, \int_{-\infty}^\infty dy \,
\mathrm{e}^{-2\kappa |y|(p+2)} \psi_0^2 + \sum_{\{n\}} \, \int \frac{dm}{\kappa}
f_{4D}(m_4) \, \int_{-\infty}^\infty dy \, \mathrm{e}^{-2\kappa
|y|(p+2)} \psi_m^2(y).
\end{equation}
But by virtue of the relations (\ref{OrtogonalityRelationsIIP}) and
(\ref{RSIIp-psi-mass}), the dependence on the $y$ coordinate drops
out completely obtaining
\begin{equation}
f= \sum_{\{n\}} \, f_{4D}(m_4|_{m=0}) + \sum_{\{n\}} \,
\int_0^\infty \frac{dm}{\kappa} f_{4D}(m_4).
\end{equation}
This expression coincides exactly with the one obtained by the zeta
function method (\ref{forceRSIIfinalshort}) once one
performs the trick (\ref{trick}).

Before ending this section, some comments are in order. Notice that
there is no factor depending on the number of compact dimensions $p$
in front of the effective 4D Casimir force, as it was reported in
\cite{Linares:2007yz,Linares:2008am}. Also, because the integration
in the non-compact extra dimension has been carried out, it is not
longer necessary to evaluate the eigenfunctions $\psi_m(y)$ on the
brane. In fact the contribution of the whole size of the plates in
the $y$ direction has been considered.

\section{Randall-Sundrum I$p$ models}\label{secRSIp}

\subsection{Mode structure}

As a final example we discuss the case of a bulk scalar field in a
RSI$p$ model. In this case the metric is given again by
(\ref{mebra}), but this time the coordinate $y$ is compact ($0\leq
y\leq \pi r$). Thus the setup allows two (3+$p$)-branes to lie,
respectively, at $y=0,\pi r$. The Casimir effect for a massless
scalar field in this model was computed using the zeta function
method for $p=0$ in \cite{Frank:2007jb} and for
arbitrary $p$ in \cite{Frank:2008dt}. We address this problem here
for a massive scalar field of mass $\mu$. As is well known
\cite{Gherghetta:2000qt}, when the higher dimensional scalar field
is massive there does not exist a zero mode solution of the
equations of motion with simple Neumann or Dirichlet boundary
conditions. In order to overcome this problem it is necessary to
modify the boundary action and include boundary mass terms
\begin{eqnarray}\label{ec.cers}
S &=& S_{\Phi} + S_{\mathrm{Brane}} \\
S_{\Phi} &=& \frac{1}{2}\ \int d{}\,^4x\,dy\, \prod_{j=1}^p R_j \, d
\theta_j \,  \sqrt{-g} \left(g^{MN}\partial_M \Phi \, \partial_N
\Phi -\mu^2 \Phi^2\right)\,,\label{ec.cers} \\
S_{\mathrm{Brane}} &=& -\int d^4x\, dy\, \prod_{j=1}^p R_j \, d
\theta_j \,\sqrt{-g}\,2b\kappa\left[\delta(y)-\delta(y-\pi r)\right]
\Phi^2 \, ,
\end{eqnarray}
where $b$ is a dimensionless constant parametrisation the boundary
mass in units of $\kappa$. The $S_{\mathrm{Brane}}$ allows to
implement various boundary conditions corresponding to different
mode's localizations in the $y$ direction as we explain now. The
resulting field equations are, by letting
$\Phi(x,y,\theta_j)=\varphi(x)\psi(y)\prod_{j=1}^p
\Theta_j(\theta_j)$,
\begin{eqnarray}
\left(\partial_{\theta_j}^2+m_{\theta_j}^2 R_j^2
\right)\Theta_j\left(\theta_j\right)&=&0\\
\left(\partial_{y}^{2}-(4+p)\kappa\,sgn(y)\partial_{y}+m^2e^{2\kappa|y|}-\mu^2
- b\kappa(\delta(y)-\delta(y-\pi r))\right)\psi(y) &=& 0
\label{eqydelta}\\
\left(\Box_4+m_4^2\right)\phi(x)&=&0
\end{eqnarray}
with $m_{\theta_j},m$ separations constants so that the effective
mass of the scalar field can be read as $m_4^2:=
m_{\theta_j}^2+m^2$. Eigenfunctions and eigenvalues of the $p$
coordinates $\theta_j$ are given by (\ref{Theta}) and we do not
elaborate further on them. In the $y$ direction the eigenfunctions
are accounted for by subjecting them to the modified Neumann
boundary conditions
\begin{equation}\label{Bcond}
\left[ \frac{\partial\psi}{\partial y} - b \kappa
\mathrm{sgn}(y)\psi  \right]_{y=0,\pi r} =0.
\end{equation}
If we write the higher dimensional mass in units of $\kappa$, i.e.
$\mu ^2 \equiv a \kappa^2$, the above equations depend on the two
arbitrary mass parameters: $a$ and $b$. For generic values of these
parameters  there are not solutions to the boundary conditions,
however if $b=\alpha\pm\gamma$ where  $\alpha = \frac{4+p}{2}$ and
$\gamma \equiv \sqrt{\alpha^2+a}$, a zero mode solution exists.
Assuming $\gamma$ to be real, the only free parameter has a range $-
\infty < b < \infty$ and using it, the scalar zero mode can be
localized anywhere in the bulk
\begin{equation}\label{ecmcn}
\psi_0(y)=\begin{cases}
\sqrt{\frac{(b-[\alpha-1])\kappa}{\left(e^{2(b-[\alpha-1])\kappa\pi
r}-1\right)}}e^{(b-[\alpha-1])\kappa y}, &b-[\alpha-1]>0
\quad\text{localized in IR brane}\\
\frac{1}{\sqrt{2\pi r}}, &b-[\alpha-1]=0 \quad\text{no
localization}\\
\sqrt{\frac{\left|b-[\alpha-1]\right|\kappa}{\left(1-e^{-2\left|b-[\alpha-1]\right|\kappa\pi
r}\right)}}e^{(b-[\alpha-1])\kappa y}, &b-[\alpha-1]<0
\quad\text{localized in UV brane}
\end{cases}
\end{equation}
which satisfies the normalization condition
\begin{equation} \label{RSIp-orto-0}
\int_{0}^{\pi r}dy \,e^{-(2+p)\kappa|y|}\psi_{0}^2(y)=1.
\end{equation}
These results generalize the case $p=0$ ($\Rightarrow \, \alpha =
2$) \cite{Gherghetta:2000qt}.

As for the massive modes we have
\begin{equation}\label{ecscnc1}
\psi_{m}(y)=e^{\alpha\kappa y} \left[ c_1
J_{\gamma}\left(\frac{m}{\kappa}e^{\kappa y}\right)+ c_2
Y_{\gamma}\left(\frac{m}{\kappa}e^{\kappa y}\right) \right]\,.
\end{equation}
where $c_{1,2}$ are arbitrary constants and $J_{\gamma}$,
$Y_{\gamma}$ are Bessel's functions of order $\gamma$. Imposing the
boundary conditions in the low energy regime $m<<\kappa$ with {\em
large} $r$: $\kappa r>>1$ allows to obtain the approximated
Kaluza-Klein mass spectrum
\begin{equation}\label{ecmn}
\frac{m_n}{\kappa}=\left(n+\frac{\gamma}{2}-\frac{3}{4}
\right)\pi\mathrm{e}^{-\kappa\pi r}\,,\qquad n=1,2,\dots \quad.
\end{equation}
Thus we have that the 4D mass is given by
\begin{equation}\label{m4RSIp}
m_4^2 = \left \{ \begin{array}{cc} \sum_{j=1}^p \frac{n_j^2}{R_j^2},
& n=0 ,\\
\sum_{j=1}^p \frac{n_j^2}{R_j^2} +
\left(n+\frac{\gamma}{2}-\frac{3}{4} \right)^2 \kappa^2 \pi ^2
\mathrm{e}^{-2\kappa\pi r},  & n=1,2,\dots \end{array} \right.
\end{equation}
The constants in (\ref{ecscnc1}) are chosen in such a way that the
orthogonality relations are
\begin{equation}\label{RSIp-orto-n}
\int_{0}^{\pi r}dy
\,e^{-(2+p)\kappa|y|}\psi_{m}(y)\psi_{m'}(y)=\delta_{mm'}.
\end{equation}

\subsection{Zeta function approach}

Now the energy density per unit of ($p+3$) volume of the plate, which is given by  $ \left( L^2 \times \prod_{j=1}^p (2\pi R_j) \times
\frac{2}{(p+3)\kappa}(1-\mathrm{e}^{-(p+3)\kappa\pi r})  \right)$, takes the form
\begin{eqnarray}
{\mathcal E}_{RSIp} &=& \frac{1}{2} \prod_{i=1,2}
\int_{-\infty}^\infty \frac{dk_i}{2 \pi} \sum_{\{n_j\}} \left(
\sum_{N=1}^\infty \omega_{k_i,N,n_j}\left(\sum_{j=1}^{p}
\frac{n_j^2}{R_j^2} \right)- \right.  \nonumber \\
& & \left. \hspace{5.5cm} l \int_{-\infty}^\infty \frac{dk_3}{2 \pi}
\omega_{k_i,k_3,n_j}\left(\sum_{j=1}^{p} \frac{n_j^2}{R_j^2} \right)
\right) \nonumber \\
& & + \frac{1}{2} \prod_{i=1,2} \int_{-\infty}^\infty \frac{dk_i}{2
\pi} \sum_{n=1} \sum_{\{n_j\}} \left( \sum_{N=1}^\infty
\omega_{k_i,N,n_j,m}\left(\sum_{j=1}^{p} \frac{n_j^2}{R_j^2}+m_n^2
\right)- \right.  \nonumber \\
& & \left. \hspace{5.5cm} l \int_{-\infty}^\infty \frac{dk_3}{2 \pi}
\omega_{k_i,k_3,n_j,m}\left(\sum_{j=1}^{p} \frac{n_j^2}{R_j^2}+m_n^2
\right) \right) \, ,
\end{eqnarray}
where $\{n_j\}$ denotes the set $\{n_1, n_2, \dots, n_p|n_1 \in
{\mathbb{Z}}, \dots , n_p \in {\mathbb{Z}} \}$,
\begin{equation}
\omega_{k_i,N,n_j,m_n}\left(\sum_{j=1}^{p}\frac{n_j^2}{R_j^2}+m^2
\right) \equiv \sqrt{k_{1}^2+ k_{2}^2 +\left(\frac{\pi
N}{l}\right)^2+\sum_{j=1}^{p}\frac{n_j^2}{R_j^2}+m_n^2},
\end{equation}
and
\begin{equation}
\omega_{k_i,k_3,n_j,m_n}\left(\sum_{j=1}^{p}
\frac{n_j^2}{R_j^2}+m_n^2 \right) \equiv \sqrt{k_{1}^2+ k_{2}^2 +
k_3^2 + \sum_{j=1}^{p} \frac{n_j^2}{R_j^2}+m_n^2}.
\end{equation}
Next, upon integrating, we have
\begin{equation}
{\mathcal E}_{RSIp}(\mu)= \sum_{\{n_j\}} \left( {\mathcal
E}_{4D}\left(\sqrt{\sum_{j=1}^{p} \frac{n_j^2}{R_j^2}}\right) +
\sum_{n=1}^\infty {\mathcal E}_{4D}\left(\sqrt{\sum_{j=1}^{p}
\frac{n_j^2}{R^2}+m_n^2}\right)\right),
\end{equation}
where ${\mathcal E}_{4D}$ is given by (\ref{energydensresult}).
Deriving this expression with respect to the separation between
plates one gets
\begin{equation} \label{F-RSIp-zeta}
f_{RSIp}(\mu)= \sum_{\{n_j\}} \left(
f_{4D}\left(\sqrt{\sum_{j=1}^{p} \frac{n_j^2}{R_j^2}}\right) +
\sum_{n=1}^\infty f_{4D}\left(\sqrt{\sum_{j=1}^{p}
\frac{n_j^2}{R^2}+m_n^2}\right)\right),
\end{equation}
As a particular case we have $\mu=p=0$ \cite{Frank:2007jb}
\begin{eqnarray}
f_{RSIp}(0) &=& f_{4D}\left( 0 \right) + \sum_{n=1}^\infty
f_{4D}\left( \left(n+\frac{1}{4} \right)\pi\mathrm{e}^{-\kappa\pi r}
\right) \\
& = & - \frac{\pi^2}{480 l^4}- \frac{\mathrm{e}^{-2\kappa\pi r}}{8}
\sum_{n=1}^\infty \left(n+\frac{1}{4} \right)^2 \left[\frac{3}{l^2}
\sum_{N=1}^{\infty}\frac{1}{N^2}K_{2}\left( 2 Nl \left(n+\frac{1}{4}
\right)\pi\mathrm{e}^{-\kappa\pi r}\right) \right. \nonumber \\
& & \left. + \frac{2}{l} \left(n+\frac{1}{4}
\right)\pi\mathrm{e}^{-\kappa\pi r} \sum_{N=1}^{\infty}
\frac{1}{N}K_{1}\left(2 Nl \left(n+\frac{1}{4}
\right)\pi\mathrm{e}^{-\kappa\pi r}\right)\right]. \nonumber
\end{eqnarray}

\subsection{Green's function approach}

Once again the calculation using the Green's function
method relies on the fact that the $(5+p)$D Greens function can be
written in terms of the 4D Green's function via
\begin{eqnarray}
G_{(5+p)D}(x,y,\theta;x',y',\theta ') &=& \sum_{\{n_j \}}
\prod_{j=1}^p \Theta_{n_j}^*(\theta_j) \Theta_{n_j}(\theta'_j)
\psi_0(y) \psi_0(y') G_{4D}(x,x'; m_4|_{m=0}) \nonumber \\
&+& \sum_{\{n_j \}} \sum_{n=1}^\infty \prod_{j=1}^p
\Theta_{n_j}^*(\theta_j) \Theta_{n_j}(\theta'_j) \psi_n(y)
\psi_n(y') G_{4D}(x,x'; m_4),
\end{eqnarray}
where $m_4$ is given by (\ref{m4RSIp}). Introducing this expression
in (\ref{t2}) and then in (\ref{ForceT}) we obtain
\begin{equation} \label{f-RSIp-G}
f= \sum_{\{n_j\}} \, f_{4D}(m_4|_{m=0}) \, \int_{0}^{\pi r} dy
\, \mathrm{e}^{-2\kappa |y|(p+2)} \psi_0^2 + \sum_{\{n_j\}} \,
\sum_{n=1}^\infty f_{4D}(m_4) \, \int_{0}^{\pi r}dy \,
\mathrm{e}^{-2\kappa |y|(p+2)} \psi_m^2(y).
\end{equation}
And now, by virtue of the orthogonality relations, Eqs. (\ref{RSIp-orto-0}) and (\ref{RSIp-orto-n}), the dependence on the $y$
coordinate drops out to yield
\begin{equation}
f= \sum_{\{n\}} \, f_{4D}(m_4(\{n_j\}, n=0)) + \sum_{\{n\}} \,
\sum_{n=1} f_{4D}(m_4(\{n_j\},n)).
\end{equation}
This expression coincides exactly with the one obtained by the zeta
function regularization method, Eq. (\ref{F-RSIp-zeta}).

\section{Discussion}\label{conclusions}

The old idea that our world is embedded in a spacetime with
dimension higher than four has reemerged in brane world models which
have revealed windows to look for deviations from standard physics
mostly in high energy physics \cite{Allanach:2004ub,Csaki:2004ay}
and cosmology (e.g.
\cite{Maartens:2003tw,Elizalde:2003bz,Elizalde:2006iu,Maartens:2010ar}).
Nevertheless low energy tests may also provide some insight into
possible imprints of extra dimensions including in particular the
Casimir force
\cite{Poppenhaeger:2003es,Linares:2005cj,Frank:2007jb,Linares:2007yz,Linares:2008am,
Frank:2008dt,Saharian:2008gs,Teo:2008ah,Teo:2009tm,Elizalde:2009nt,
Cheng:2009xf,Teo:2009dd,Cheng:2009bv,Kharlanov:2009pv,Cheng:2009uu,
Teo:2009bv,Bellucci:2009hh,Rypestol:2009pe,Nouicer:2009ze,Beneventano:2010wy,Elizalde:2010av}.
Such force is sensitive to the mode structure of the field which in
turn depends on the features of the background spacetime and bounds
can be set for the values of the parameters of given brane world
models which produce Casimir forces deviating from known data beyond
the corresponding uncertainties.

To determine the Casimir force one can made use of either of two
well known approaches: Green's function and Zeta function. In the
case of flat spacetimes both approaches yield the same result (See
eg. \cite{Milton:2001yy}), however, for brane worlds one usually
assumes this is the case. In this work we have actually shown
Green's function and Zeta function yield the same Casimir force for
the case of Universal Extra Dimensions and Randall-Sundrum models
with one or two branes added by $p$ compact dimensions. These
results correct in particular an erroneous difference between the
Casimir force obtained by Green's function technique
\cite{Linares:2007yz,Linares:2008am} and the one obtained from zeta
function \cite{Frank:2007jb,Frank:2008dt} for a massless scalar
field in the case of a single brane Randall-Sundrum scenario added
by $p$ compact dimensions.  The origin of the difference in this
case was due to an incorrect hyper-volume for the plates subject to
the Casimir force in \cite{Linares:2007yz,Linares:2008am}.

The coincidence of the above two approaches to the Casimir force can
be understood as follows. Although the Green's function technique
involves further details of the mode decomposition of the
corresponding fields, it is due to their orthogonality relations
which involve the correct hyper-volume factor of the plates that one
can literally eliminate the mode eigenfunctions from the Casimir
force. This is neatly seen in the case of UXD, Eq. (\ref{f-UEXD-G}).
This holds similarly for the cases of RSII-$p$ and RSI-$p$, as can
be seen from Eqs. (\ref{f-RSIIp-G}) and (\ref{f-RSIp-G}),
respectively.

Hence given the equivalence of the zeta function and Green's
function to determine the Casimir force in Randall-Sundrum and
Universal Extra Dimensions models one can conclude that localization
of the field modes does not play a role as far as the Casimir force
is concerned. This is so due to the fact that zeta function is
actually insensitive to the form of the mode eigenfunctions but only
to dispersion relations. On the side of the Green's function
approach, while built explicitly on an eigenfunction expansion, it
looses them by virtue of the orthogonality relations they fulfill
which just absorbs the hyper-volume of the plates. This is
reassuring since it has been noticed that localizing all the fields
of the Standard Model  to the brane located at $y=\pi R$ leads to
problems with the phenomenology of proton decay, Flavor Changing
Neutral Currents (FCNC) effects and neutrino masses
\cite{Gherghetta:2000qt}. Indeed, although originally the $p$ extra
compact dimensions were added to the Randall-Sundrum models to
produce localization of gauge fields, such a feature seems not to be
required anymore.

It should be stressed an important pattern has put forward in this
work for the resulting Casimir force in brane worlds, namely, it can
be expressed as the sum of two terms each one having the specific
form of the 4D Casimir force: the first one containing in particular
the zero mode defined by the extra dimensions and the second one
including the full Kaluza-Klein tower of massive modes. This feature
simplified importantly the analysis and in particular allowed to
adopt the full machinery of previous results in $d+2$ Minkowski
spacetime \cite{Ambjorn:1981xw}.

It would be rather interesting to study other brane world models to
test the equivalence we have here proved between the zeta function
and Green's function approaches in the calculation of the Casimir
force. Indeed, it should be possible to have a general proof of it
at least for sufficiently symmetric spacetimes
\cite{Birrell:1982ix}.

\section{Appendix: Casimir force in d+2 Minkowski spacetime}\label{apendice}

In this appendix we compute the Casimir force for a massive scalar
field of mass $\mu$ in $(d+2)$-dimensional Minkowski spacetime. We
discuss the calculation using the two methods we are interested in:
the {\it zeta function method} and the {\it Green's function
method}. Although for a given mass $\mu \neq 0$ the force can be
computed at the very end only numerically, it is possible to give an
analytical expression of it for a generic mass. The aim to show this
computation is twofold: i) to avoid unnecessary repetitions of the
calculation throughout the paper and ii) to allow us comparison with
recent results in the literature.

\subsection{Zeta function approach}

This computation was originally discussed in \cite{Ambjorn:1981xw}
and we adapt it here to our notation. Our starting point is equation
(\ref{intenergydimreg})
\begin{equation}\label{intenergydimreg2}
{\mathcal E} = \frac{E_{plates}- E_0}{L^2} = \frac{1}{2}
\prod_{i=1,2} \int_{-\infty}^\infty \frac{dk_i}{2 \pi} \left(
\sum_{N=1}^\infty \omega_{k_1,k_2,N}(\mu) - l \int_{-\infty}^\infty
\frac{dk_3}{2 \pi} \omega_{k_1,k_2,k_3}(\mu) \right).
\end{equation}
In terms of the integral
\begin{equation}\label{basicint}
I_d(\alpha^2) \equiv \frac12 \int \frac{d^dk}{(2\pi)^d} \sqrt{k_d^2
+ \alpha^2},
\end{equation}
we can rewrite equation (\ref{intenergydimreg2}) as
\begin{equation}
{\mathcal E} =  \sum_{N=1}^\infty I_{d=2} \left( \frac{N^2
\pi^2}{L^2}+\mu^2\right)- l\, I_{d=3}(\mu^2).
\end{equation}
In order to be general, we shall evaluate the Casimir energy for a
scalar field in D = $d+2$ dimensional Minkowski spacetime, between
hyperplanes of dimension $d$. This problem is known as the Casimir
effect of codimension one \cite{Ambjorn:1981xw}
\begin{equation}\label{intenergyDdim}
{\mathcal E}_{d+2} = \frac{E_{plates}-E_0}{L^d} =  \sum_{N=1}^\infty
I_{d} \left( \frac{N^2 \pi^2}{l^2}+\mu^2\right)- l \,
I_{d+1}(\mu^2).
\end{equation}
Using the Euler representation for the gamma function
\begin{equation}\label{gammafun}
\Gamma(z)=g^z \int_0^\infty e^{-gt}t^{z-1}dt,
\end{equation}
the integral (\ref{basicint}) can be rewritten employing the
Schwinger proper time representation for the square root as
\begin{equation}
I_d(\alpha^2)= \frac{1}{2 \Gamma(-1/2)} \int \frac{d^dk}{(2\pi)^d}
\int_0^\infty \frac{dt}{t} t^{-1/2} e^{-t (k_d^2 + \alpha^2)}.
\end{equation}
Performing the Gaussian integral first and using (\ref{gammafun})
again we have
\begin{equation}
I_d(\alpha^2)= -\frac{1}{2} \frac{1}{(4\pi)^{\frac{d+1}{2}}}
\Gamma\left(-\frac{d+1}{2} \right) \alpha^{d+1},
\end{equation}
where we have used the value $\Gamma \left(-\frac12
\right)=-2\sqrt{\pi}$. Substituting this result in
(\ref{intenergyDdim}) we obtain
\begin{equation}\label{energyint}
{\mathcal E}_{d+2} = -\frac{1}{2(4\pi)^{\frac{d+1}{2}}} \left(
\Gamma \left(-\frac{d+1}{2}\right) \left( \frac{\pi}{l}
\right)^{d+1} \sum_{N=1}^\infty \left(N^2+ \frac{
l^2\mu^2}{\pi^2}\right)^{\frac{d+1}{2}}+ \frac{\Gamma
\left(-\frac{d+2}{2}\right)}{\Gamma \left(-\frac{1}{2}\right)} \, l
\, \mu^{d+2}\right).
\end{equation}
In order to compute the sum we use the Epstein-Hurwitz function
which is defined as
\begin{eqnarray}\label{Epstein}
\zeta_{EH}(s,a^2)=\sum_{N=1}^\infty (N^2+a^2)^{-s}
&=&-\frac{(a^2)^{-s}}{2}+ \frac{\sqrt{\pi}\,
\Gamma\left(s-\frac{1}{2}\right)}{2\Gamma\left(s\right)}(a^2)^{\frac{1}{2}-s}
\nonumber\\
&&+\frac{2\pi^s}{\Gamma\left(s\right)}(a^2)^{-\frac{s}{2}+\frac{1}{4}}
\sum_{n=1}^{\infty}n^{s-\frac{1}{2}}K_{s-\frac{1}{2}}\left(2\pi
n\sqrt{a^2}\right),
\end{eqnarray}
where $K$ is the modified Bessel function of second type. In our
case $s=-\frac{d+1}{2}$ and $a=\frac{l \mu}{\pi}$. It turns out that
the second term in (\ref{Epstein}) cancels with the second term in
(\ref{energyint}). Often this cancelation is not performed
explicitly but by an equivalent argument the second term in the
Epstein-Hurwitz function is discarded \cite{Ambjorn:1981xw}. This is
why people claim that in the zeta function regularization method it
is not necessary to subtract any quantity and that a finite result
comes directly considering only the first integral in
(\ref{intenergydimreg2}). The final expression for the energy is
\begin{equation}\label{EnergyGenRes}
{\mathcal E}_{d+2} = -\frac{1}{2(4\pi)^{\frac{d+1}{2}}} \left(
-\frac12 \Gamma \left(-\frac{d+1}{2}\right)  \mu^{d+1} +
\frac{2}{\sqrt{\pi}} \frac{\mu^{\frac{d+2}{2}}}{l^{\frac{d}{2}}}
\sum_{n=1}^\infty \frac{1}{n^{\frac{d+2}{2}}}K_{-\frac{d+2}{2}}
(2nl\mu) \right).
\end{equation}
The first term is a constant energy and therefore we discard it
because does not contribute to the Casimir force. Deriving the
energy with respect to the separation $l$ between the hyperplanes,
we obtain the ($d$+2)-dimensional Casimir force
\begin{equation}
f_{d+2}=-\frac{d {\mathcal E}_{d+2}}{d l} =
\frac{2}{(4\pi)^{\frac{d+2}{2}}} \mu^{\frac{d+2}{2}} \frac{d}{d l}
\left[ \frac{1}{l^{\frac{d}{2}}} \sum_{n=1}^\infty
\frac{1}{n^{\frac{d+2}{2}}}K_{-\frac{d+2}{2}} (2nl\mu) \right],
\end{equation}
which can be evaluated explicitly. Using the properties
$K_{-\nu}(z)=K_\nu(z)$ y $z\partial_z K_{\nu}(z)=-z K_{\nu-1}(z)-\nu
K_{\nu }(z)$ we obtain finally
\begin{equation}\label{CasimirGenRes}
f_{(d+2)}(\mu)= -2 \left
(\frac{\mu}{4\pi}\right)^{\frac{d+2}{2}}\left[\frac{1}{l^{\frac{d+2}{2}}}
\sum_{n=1}^{\infty}\frac{1}{n^{\frac{d+2}{2}}}K_{\frac{d+2}{2}}(2
nl\mu) - \frac{2 \mu}{l^{\frac{d}{2}}}\sum_{n=1}^{\infty}
\frac{1}{n^{\frac{d}{2}}}K_{\frac{d+4}{2}}(2 nl\mu)\right].
\end{equation}
In particular we are interested in the 4D Casimir force, which is
obtained setting $d=2$ in the formula above
\begin{equation}
f_{4D}(\mu)= -\frac{\mu^2}{8\pi^2} \left[\frac{1}{l^2}
\sum_{n=1}^{\infty}\frac{1}{n^2} K_2(2 nl\mu) - \frac{2\mu}{l}
\sum_{n=1}^{\infty}\frac{1}{n} K_3(2 nl\mu) \right].
\end{equation}
Sometimes this expression is presented in a slightly different way,
which can be obtained by using the identity
$K_\nu(z)=K_{\nu-2}(z)+\frac{2(\nu - 1)}{z} K_{\nu - 1}(z)$
\begin{equation}
f_{4D}(\mu)= \frac{\mu^2}{8\pi^2} \left[\frac{3}{l^2}
\sum_{n=1}^{\infty}\frac{1}{n^2} K_2(2 nl\mu) + \frac{2\mu}{l}
\sum_{n=1}^{\infty}\frac{1}{n} K_1(2 nl\mu) \right].
\end{equation}

\subsection{Green's function approach}

Our starting point is the integral (\ref{eq:f4Dmu0}), but in the spirit
of generality, in analogy with the zeta function
method we allow to have $d$ transverse dimensions, namely
\begin{equation}
f_{(d+2)}(\mu)=\frac{1}{2}\int \frac{d^dk}{(2\pi)^d}\int
\frac{d\xi}{2\pi} \left( \frac{2}{l}
\sum_{k=1}^{\infty}\frac{\frac{\pi^2k^2}{l^2}}{\rho^2+\mu^2+
\frac{\pi^2k^2}{l^2}}+ \sqrt{\rho^2+\mu^2}\right).
\end{equation}
This integral can be performed straightforward in polar coordinates,
in terms of the volume of the unitary $d$-dimensional sphere that we
denote by $vol(S^d)$
\begin{equation}
\int \frac{d^d k}{(2 \pi)^d} \int \frac{d \xi}{2\pi} \rightarrow
\frac{vol(S^{d})}{(2\pi)^{d+1}}\int_0^\infty \rho^d d \rho,
\hspace{0.5cm}\mbox{with,}\hspace{0.5cm}
vol(S^{d})=\frac{2\pi^{\frac{d+1}{2}}}{\Gamma\left(\frac{d+1}{2}\right)}.
\end{equation}
In polar coordinates, Eq. (5) is rewritten as
\begin{eqnarray}
f_{(d+2)}(\mu)&=&\frac{vol(S^{d})}{2(2\pi)^{d+1}}\left[ \frac{2}{l}
\sum_{k=1}^{\infty}\frac{\pi^2k^2}{l^2}\int_0^\infty d\rho
\frac{\rho^d}{\rho^2+\mu^2+ \frac{\pi^2k^2}{l^2}} + \int_0^\infty
d\rho \, \rho^d \sqrt{\rho^2+\mu^2}\right] \label{intintgreen}\\
&=&
\frac{vol(S^{d})\Gamma\left(\frac{d+1}{2}\right)}{4(2\pi)^{d+1}}\left[
\frac{2 \, \Gamma\left(\frac{1-d}{2}\right)}{l}\sum_{k=1}^\infty
\frac{\pi^2k^2}{l^2}
\left(\mu^2+\frac{\pi^2k^2}{l^2}\right)^{\frac{d-1}{2}} +
\frac{\Gamma\left(-\frac{d+2}{2}\right)}{\Gamma\left(-\frac{1}{2}\right)}\mu^{d+2}
\right],\nonumber
\end{eqnarray}
where we have used the result
\begin{equation}
\int_0^\infty d \rho \, \frac{\rho^d}{\left( \rho^2 + c
\right)^s}=\frac{1}{2} \, \Gamma\left(\frac{d+1}{2}\right)\,
\frac{\Gamma\left(s-\frac{1}{2}-\frac{d}{2}\right)}
{\Gamma\left(s\right)} \, c^{\frac{d}{2}+\frac{1}{2}-s},
\end{equation}
with $s=1$ and $c= \mu^2 +\frac{ \pi^2 k^2}{l^2}$ in the first
integral, and $s=-1/2$ and $c=\mu^2$ in the second. The next step in
the computation is to notice that the two terms together in
(\ref{intintgreen}) can be rewritten in terms of the derivative of a
Epstein-Hurwitz function (\ref{Epstein}). In order to show this
consider the infinite series in the first term which we shall denote
as $\cal S$
\begin{equation}
{\cal S}=\frac{1}{l}\sum_{k=1}^\infty \frac{\pi^2k^2}{l^2}
\left(\mu^2+\frac{\pi^2k^2}{l^2}\right)^{\frac{d-1}{2}}=-\frac{1}{d+1}\left(
\frac{\pi}{l}\right)^{d+1}\frac{d}{dl} \sum_{k=1}^\infty
\left(k^2+\frac{l^2\mu^2}{\pi^2}\right)^{\frac{d-1}{2}},
\end{equation}
but the series written in this way is precisely the Epstein-Hurwitz
zeta function (\ref{Epstein}). In terms of it
\begin{equation}
{\cal S}=-\frac{1}{d+1}\left(
\frac{\pi}{l}\right)^{d+1}\frac{d}{dl}\,
\zeta_{EH}\left(-\frac{d+1}{2},\left( \frac{l\mu}{\pi}
\right)^2\right),
\end{equation}
and computing explicitly the derivative we obtain
\begin{equation}
{\cal S}=-\frac{1}{d+1}\frac{1}{\Gamma\left(\frac{d+1}{2}\right)}
\left[ \frac{\Gamma\left(-\frac{d+2}{2}\right)}{2 \sqrt \pi}
\mu^{d+2}+\frac{2}{\sqrt \pi}
\mu^{\frac{d+2}{2}}\frac{d}{dl}\left(l^{-\frac{d}{2}}
\sum_{n=1}^{\infty}n^{-\frac{d+2}{2}}K_{-\frac{d+2}{2}}\left(2 nl\mu
\right) \right)\right].
\end{equation}
Using the identity
$-(d+1)\Gamma\left(-\frac{d+1}{2}\right)=2\Gamma\left(-\frac{d-1}{2}\right)$,
and since $\Gamma \left( -\frac{1}{2}\right)=-2 \sqrt{\pi}$, we
obtain finally
\begin{equation}
2{\cal S}\Gamma\left(\frac{1-d}{2}\right) +
\frac{\Gamma\left(-\frac{d+2}{2}\right)}{\Gamma \left(
-\frac{1}{2}\right)} \mu^{d+2}=\frac{2}{\sqrt \pi}
\mu^{\frac{d+2}{2}}\frac{d}{dl}\left(l^{-\frac{d}{2}}
\sum_{n=1}^{\infty}n^{-\frac{d+2}{2}}K_{-\frac{d+2}{2}}\left(2 nl\mu
\right) \right).
\end{equation}
But these are precisely the terms inside the brackets in equation
(\ref{intintgreen}), so we get the result
\begin{equation}
f_{d+2}(\mu)= \frac{vol(S^{d})\Gamma\left(\frac{d+1}{2}\right)\sqrt
\pi}{(2\pi)^{d+2}} \mu^{\frac{d+2}{2}}
\frac{d}{dl}\left(l^{-\frac{d}{2}}
\sum_{n=1}^{\infty}n^{-\frac{d+2}{2}}K_{-\frac{d+2}{2}}\left(2 nl\mu
\right) \right).
\end{equation}
We can evaluate explicitly the derivative. Using the properties
$K_{-\nu}(z)=K_\nu(z)$ y $z\partial_z K_{\nu}(z)=-z K_{\nu-1}(z)-\nu
K_{\nu }(z)$ and substituting the value of $vol(S^d)$ we obtain
finally
\begin{equation}
f_{(d+2)}(\mu)= -2 \left
(\frac{\mu}{4\pi}\right)^{\frac{d+2}{2}}\left[\frac{1}{l^{\frac{d+2}{2}}}
\sum_{n=1}^{\infty}\frac{1}{n^{\frac{d+2}{2}}}K_{\frac{d+2}{2}}(2
nl\mu) - \frac{2 \mu}{l^{\frac{d}{2}}}\sum_{n=1}^{\infty}
\frac{1}{n^{\frac{d}{2}}}K_{\frac{d+4}{2}}(2 nl\mu)\right].
\end{equation}
This expression of the force coincides exactly with the one obtained
above using the zeta function approach.

\begin{acknowledgments}
This work was partially supported by Mexico's National Council of
Science and Technology (CONACyT), under grants
CONACyT-SEP-2004-C01-47597 and CONACyT-SEP-2005-C01-51132F. O. P. is
partially supported by grant SIINV-UNACH-2009-01-SYM-SNV-146-09 and
PROMEP.
\end{acknowledgments}

\bibliography{bibliography}

\end{document}